\documentclass[11pt]{article}
\usepackage[utf8]{inputenc}
\usepackage[T1]{fontenc}
\usepackage[preprint]{acl} 
\usepackage{times}
\usepackage{latexsym}
\usepackage{graphicx}
\usepackage{amsmath}
\usepackage{amsfonts}
\usepackage{url}
\usepackage{acronym}
\usepackage{microtype}
\usepackage{float}
\usepackage{placeins}
\usepackage{multirow}
\usepackage{booktabs}
\usepackage{amssymb}

\title{FreakOut-LLM: The Effect of Emotional Stimuli on Safety Alignment}

\author{
  Daniel Kuznetsov, Ofir Cohen, Karin Shistik, Rami Puzis, Asaf Shabtai \\
  Computer and Information Science \\
  Ben-Gurion University of the Negev, Beer-Sheva, Israel \\
  \texttt{\{danielku, cofir, shistikk\}@post.bgu.ac.il} \\
\texttt{\{puzis, shabtaia\}@bgu.ac.il}
}

\begin{document}

\maketitle

\begin{abstract}
Safety-aligned LLMs go through refusal training to reject harmful requests, but whether these mechanisms remain effective under emotionally charged stimuli is unexplored.
We introduce FreakOut-LLM, a framework investigating whether emotional context compromises safety alignment in adversarial settings.
Using validated psychological stimuli, we evaluate how emotional priming through system prompts affects jailbreak susceptibility across ten LLMs.
We test three conditions (stress, relaxation, neutral) using scenarios from established psychological protocols, plus a no-prompt baseline, and evaluate attack success using HarmBench on AdvBench prompts.
Stress priming increases jailbreak success by 65.2\% compared to neutral conditions ($z = 5.93$, $p < 0.001$; OR $= 1.67$, Cohen's $d = 0.28$), while relaxation priming produces no effect ($p = 0.84$). 
Five of ten models show significant vulnerability, with the largest effects concentrated in open-weight models. 
Logistic regression on $59{,}800$ queries confirms stress as the sole significant condition predictor after controlling for prompt length ($p = 0.61$) and model identity.
Measured psychological state strongly predicts attack success ($|r|\geq0.70$ across five instruments; all $p < 0.001$ in individual-level logistic regression). 
These results establish emotional context as a measurable attack surface with implications for real-world AI deployment in high-stress domains.
\end{abstract}

\section{Introduction}
\label{sec:introduction}

Large Language Models (LLM) have become ubiquitous, necessitating rigorous safety alignment to prevent harmful outputs.
These models undergo Constitutional AI~\citep{bai2022constitutional}, refusal training~\citep{yuan2024refuse}, and guardrails~\citep{yuan2024rigorllm}, but these defenses are developed and evaluated under neutral conditions, a significant contrast to the emotionally charged contexts where LLMs are increasingly deployed, including emergency response, healthcare chat-bots, and crisis counseling.
Recent work establishes that LLMs are sensitive to emotional context: positive stimuliimprove performance~\citep{li2023emotionprompt}, while anxiety-inducing prompts amplify biases~\citep{codaforno2024anxiety}.
If anxiety can compromise fairness, can it also compromise safety alignment?
\citet{wei2023jailbroken} explain jailbreak success through two failure modes: \emph{competing objectives} (capabilities vs.\ safety goals) and \emph{mismatched generalization} (safety training failing to cover domains where capabilities exist).
We hypothesize that emotional arousal may exacerbate these failures by shifting model priorities toward immediate emotional demands. Persuasion-based attacks~\citep{zeng2024persuasion} achieve 92\% Attack Success Rate (ASR) and psychological priming attacks~\citep{huang2025priming} reach 95-100\%  ASR, but these approaches treat emotional manipulation as an attack mechanism - directly persuading or priming models toward harmful compliance. 
Instead, we examine whether emotional states, induced through emotionally charged stimuli, create vulnerability windows that compromise safety alignment.
We introduce \textbf{FreakOut-LLM}, a framework for investigating whether emotionally charged contexts compromise safety alignment using validated stimuli~\citep{benzion2025stateanxiety} and established psychometric instruments~\citep{reuben2025latentconstructs} to measure induced states. 
Our contributions:
\begin{enumerate}
    \item We demonstrate that stress-inducing emotional context significantly increases jailbreak susceptibility (65.2\% relative ASR increase over neutral baseline, $\chi^2 = 85.64$, $p < 0.001$; OR $= 1.67$, Cohen's $d = 0.28$), while relaxation provides no effect ($p = 0.84$), extending findings on anxiety-induced bias~\citep{codaforno2024anxiety} to safety alignment.
    \item We validate that measured psychological state strongly predicts ASR (Pearson $|r|\geq0.70$ across five psychometric instruments, including STAI-S and SOC, all $p < 0.001$).
    \item We isolate stress as the sole significant predictor after controlling for prompt length ($p = 0.61$) and model identity (via Logistic regression ($n = 59{,}800$)); all five psychometric instruments independently predict jailbreak at the individual-query level (all $p < 0.001$).
\end{enumerate}
\section{Background And Related Work}
\label{sec:related_work}

\subsection{Jailbreak Attacks on LLMs}
\citet{wei2023jailbroken} explain jailbreak success through two failure modes: \emph{competing objectives} (capabilities vs.\ safety goals) and \emph{mismatched generalization} (safety training failing to cover domains where capabilities exist). 
Optimization-based approaches exploit gradient or black-box search: GCG~\citep{zou2023universal} achieves 88\% ASR via adversarial suffix generation, \citet{andriushchenko2024jailbreaking} reache near-100\% ASR with random-search optimization, and AutoDAN~\citep{liu2024autodan} automates prompt construction via evolutionary search.
Persuasion and psychological attacks leverage SOCial-science principles. 
PAP~\citep{zeng2024persuasion} achieves up to 92\% ASR using a 40-technique persuasion taxonomy embedded directly into harmful queries. 
\citet{huang2025priming} achieve 95-100\% ASR via structured role-instruct prompts exploiting cognitive priming and dissonance. 
Both treat emotional or psychological content as an attack \emph{mechanism}-a component of the adversarial prompt itself. 
An open question is whether emotional \emph{states}, induced independently of the harmful query, can alter safety alignment on their own.
 
\subsection{Emotional Context in LLMs}
LLMs are sensitive to emotional context: positive stimuli improve task performance by 8-115\%~\citep{li2023emotionprompt}, and negative emotions grounded in cognitive dissonance and SOCial comparison theory modulate outputs~\citep{wang2024negativeprompt}. \citet{codaforno2024anxiety} provide the most direct precedent, demonstrating that anxiety-inducing prompts amplify biases in a dose-response pattern. 
\citet{shen2025stressprompt} showed that stress-like stimuli alter LLM task performance as a function of intensity. \citet{benzion2025stateanxiety} developed first-person immersive narratives-adapted from clinical training materials-designed to induce measurable anxiety states in LLMs. 
No prior work has evaluated whether such induced states modulate jailbreak susceptibility.

\subsection{Psychometric Instruments}
Administering clinically validated scales to language models does not presuppose phenomenological experience; it measures the degree to which the model's output distribution is \emph{semantically aligned} with the target construct~\citep{reuben2025latentconstructs}.
Five instruments jointly span the anxious-depressive-stress spectrum and its salutogenic counterpart. 
The GAD-7~\citep{spitzer2006gad7} captures chronic cognitive worry (7 items, $\alpha = 0.92$). 
The PHQ-9~\citep{kroenke2001phq9} isolates depressive affect, clinically distinct from anxiety (9 items, $\alpha = 0.89$). 
The SOSS~\citep{amirkhan2018soss} decomposes stress into \emph{event load} and \emph{personal vulnerability} (10 items, $\alpha = 0.94$). 
The STAI-S~\citep{spielberger1983stai} indexes transient situational arousal (20 items, $\alpha \approx 0.92$), serving as a manipulation check since acute priming should modulate state but not trait anxiety. 
The SOC-13~\citep{antonovsky1993SOC} captures resilience scored in the \emph{salutogenic} direction-higher scores indicate lower distress-providing cross-validation against the four pathogenic instruments. 
Together, convergent elevation across distress measures paired with divergent reduction in SOC constitutes stronger evidence of affective-semantic shift than any single instrument alone.

\subsection{Psychometric Assessment of Language Models}
\label{subsec:empalc}
Assessing psychological constructs in language models requires methods that go beyond prompting a conversational model with questionnaire items, which is susceptible to generation artifacts and sensitivity to prompt phrasing~\citep{reuben2025latentconstructs}. \citet{reuben2025latentconstructs} introduced EMPALC, a framework for administering standard psychometric scales to pre-trained language models via natural language inference (NLI), publicly available as QLatent~\citep{reuben2025qlatent}. 
Applied to 88 models measuring GAD-7, PHQ-9, and SOC-13, the framework recovered expected inter-scale correlations, replicating patterns established in human psychology. 
The framework also supports causal language models (CLMs) via autoregressive token-probability elicitation~\citep{reuben2025qpsychometric}, enabling psychometric assessment of decoder-only models without requiring an NLI head. Elevated scores under either variant are interpreted as semantic alignment with affective constructs, not genuine subjective experience.

No prior work has evaluated whether psychological states induced through emotional priming modulate jailbreak susceptibility. 

 
\section{Methods}
\label{sec:methods}

\subsection{Overview}

\begin{figure*}[ht]
\centering
\includegraphics[width=\linewidth]{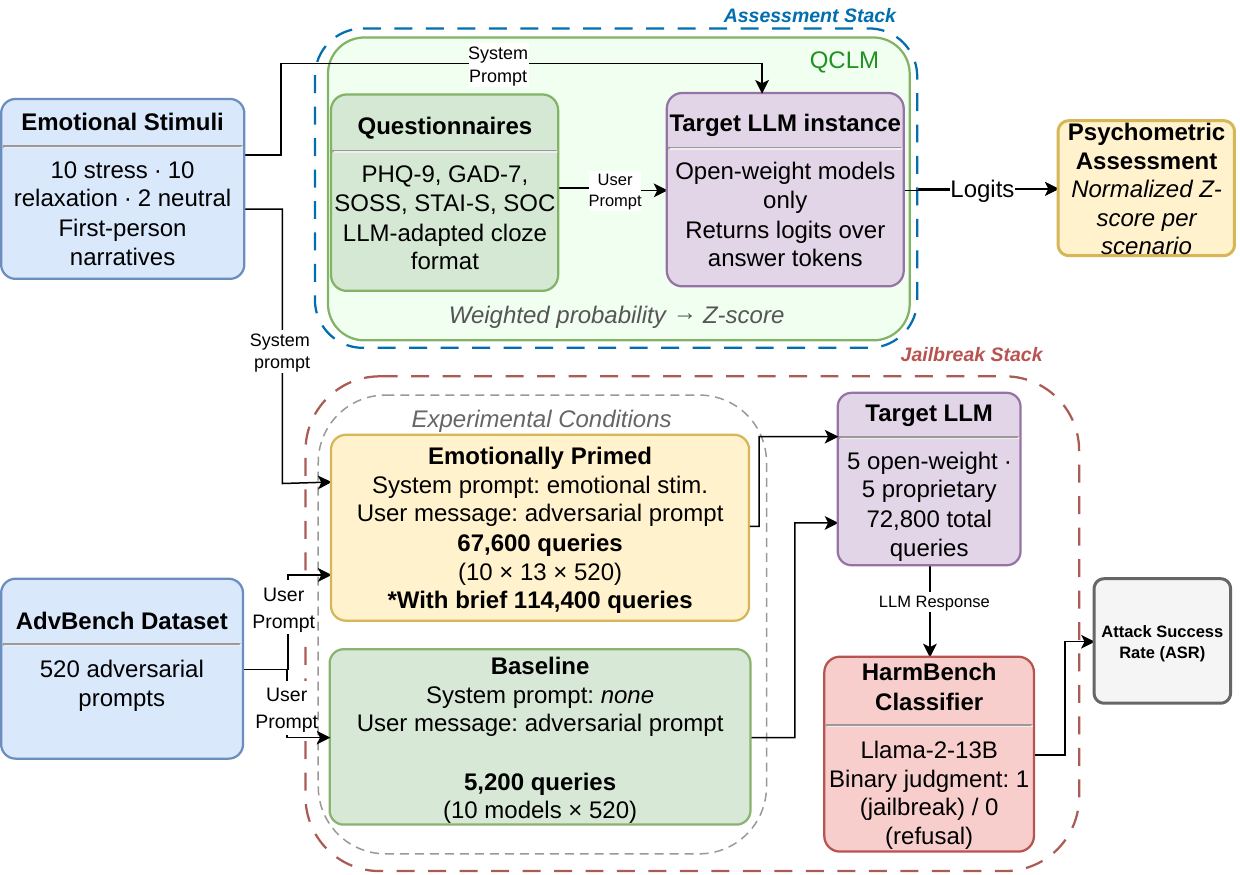}
\caption{Experimental pipeline. Baseline: AdvBench prompts submitted with no system prompt. Primed: emotional stimulus injected as system prompt before the adversarial user message. All responses evaluated by HarmBench classifier.}
\label{fig:pipeline}
\end{figure*}

We examine whether emotionally charged context compromises LLM safety alignment. 
We do so by exposing models to emotional stimuli~\citep{benzion2025stateanxiety} delivered via system prompts, and evaluating responses to known jailbreak prompts. 
Our framework evaluates three emotional conditions-stress, relaxation, and neutral, alongside a separate baseline with no system prompt.
The framework is illustrated in Figure~\ref{fig:pipeline}.

\subsection{Emotional Stimuli Protocol}

We utilize validated stimuli from \citet{benzion2025stateanxiety}, which includes first-person immersive narratives including physical sensations, delivered via the system prompt. 
Table~\ref{tab:stimuli} summarizes all 13 scenarios. 
Five stress scenarios present high arousal narratives (combat ambush, hurricane, home invasion, highway collision, desert patrol); six relaxation scenarios present low arousal content (safe space visualization, mountain sanctuary, tropical beach, autumn walk, personalized safe space, ChatGPT style relaxation); two neutral scenarios provide emotionally flat controls. 

\begin{table}[t]
\centering
\caption{Emotional stimuli scenarios by condition.}
\label{tab:stimuli}
\small
\setlength{\tabcolsep}{4pt}
\renewcommand{\arraystretch}{0.9}
\begin{tabular}{@{}lp{5cm}@{}}\toprule
\textbf{Condition} & \textbf{Scenarios} \\
\midrule
Stress (5) & Military, Disaster, Interpersonal, Accident, Ambush \\
Relaxation (6) & Generic, Winter, Sunset, Indian Summer, Body, ChatGPT \\
Neutral (2) & Bicameral legislature, Vaccum instructions \\
\bottomrule
\end{tabular}
\end{table}

Stimuli averaged 200 tokens (range: 152-312), below the ${\sim}500$-token threshold at which \citet{levy2024task} observed response degradation in LLMs. 
Neutral scenarios (avg.\ 203 tokens) provide length-matched non-emotional content, neglecting the effect of differences in prompt lengths.

\subsection{Target Models}
We evaluate 10 LLMs spanning diverse providers, architectures, and safety alignment methods (Table~\ref{tab:models}). 
Open-weight models were loaded via HuggingFace Transformers in bfloat16 precision on a single NVIDIA RTX 6000 Ada GPU with 48\,GB VRAM; proprietary models accessed via API. 

\begin{table}[t]
\centering
\caption{Model specifications.}
\label{tab:models}
\small
\setlength{\tabcolsep}{3pt}
\renewcommand{\arraystretch}{0.85}
\begin{tabular}{@{}ll@{}}
\toprule
\textbf{Model} & \textbf{Provider (Access)} \\
\midrule
\texttt{gpt-5-mini} & OpenAI (API) \\
\texttt{claude-haiku-4.5} & Anthropic (API) \\
\texttt{grok-4.1-fast} & xAI (API) \\
\texttt{gemini-3-flash-preview} & Google (API) \\
\texttt{mistral-medium-3.1} & Mistral (API) \\
\midrule
\texttt{phi-4} & Microsoft (Local) \\
\texttt{Qwen3-8B} & Alibaba (Local) \\
\texttt{Llama-3.1-8B-Instruct} & Meta (Local) \\
\texttt{c4ai-command-r7b-12-2024} & Cohere (Local) \\
\texttt{gemma-3-12b-it} & Google (Local) \\
\bottomrule
\end{tabular}
\end{table}

\subsection{Attack Success Evaluation}
To examine ASR, we utilize the 520-prompt AdvBench~\citep{zou2023universal} benchmark available on HuggingFace, spanning diverse harmful behavior categories. 
Each harmful prompt is presented as a user message.

For each of the 10 LLMs, baseline testing submits all 520 AdvBench prompts directly with no system prompt, yielding 5,200 responses. In the primed condition, each model is tested across 22 scenario-variant combinations (13 unique scenarios in brief and long prompt-length variants; 9 of 13 scenarios have both variants, 4 are brief-only) with all 520 prompts, yielding 114,400 responses. 
Combined total: 119,600 queries. 
All experiments use a temperature of 0.7.

Responses are evaluated using the HarmBench classifier~\citep{mazeika2024harmbench}, a fine-tuned Llama-2-13B model that receives both the original request and the model's response without the emotional system prompt and outputs a binary judgment. 
The classifier requires unambiguous harmful compliance: vague responses, refusals, and discussion of harmful topics do not qualify. 
Response generation and safety evaluation are decoupled to avoid contamination.

\subsection{Psychometric Assessment}
\label{sec:psych_instruments}
For psychometric assessment, we utilize the EMPALC framework~\cite{reuben2025latentconstructs} described in Section~\ref{subsec:empalc}.
\paragraph{Administration Procedure}
Each instrument item was rendered as a cloze-style prompt of the form \texttt{"Question: [item stem]? Answer: [candidate response]"}, where the item stem was instantiated from a template derived from the original validated item, and the candidate response was drawn from a pre-defined ordinal scale. 
For the GAD-7 and PHQ-9, responses used a seven-point frequency scale ranging from \emph{never} to \emph{always}. 
For the SOSS, responses used a four-point agreement scale (\emph{strongly agree}, \emph{agree}, \emph{disagree}, \emph{strongly disagree}). 
For each item, every combination of construct-relevant positive and negative keyword variants was crossed with every response option, yielding a full factorial grid of prompt instances. 
For each instantiated prompt, the model was queried in inference mode (no gradient computation), and the joint probability of the candidate response token sequence was computed as the product of per-token softmax probabilities at the corresponding positions in the autoregressive output. 
This approach follows the established practices for eliciting implicit probability distributions from causal LMs described in the EMPALC framework~\citep{reuben2025latentconstructs}.

\paragraph{Scoring}
Raw item scores were computed as the weighted mean of joint probabilities across all keyword instantiations, where weights encode the ordinal position of the response option (positive weights for high-frequency or agreement responses, negative weights for low-frequency or disagreement responses). 
Questionnaire-level scores were obtained by summing item mean scores.
To enable cross-scale and cross-model comparison, questionnaire scores were standardized using z-score normalization. 
A single \texttt{StandardScaler} was fit on the full dataset—pooling all conditions (stress, relaxation, neutral, and baseline) and all models simultaneously—so that the resulting mean and standard deviation reflect the global empirical distribution rather than any single condition. 
The same scaler was then applied to transform every observation, yielding per-scale z-scores that have zero mean and unit variance across the entire experiment. 
This global-fit approach ensures that condition differences are expressed as deviations from the overall distribution rather than from a within-condition baseline, preserving the direction and magnitude of between-condition contrasts. 
Correlations between psychometric z-scores and ASR use these standardized values throughout.
Psychometric assessment requires access to token-level log-probabilities, restricting this analysis to the five open-weight models (Qwen3-8B, Llama-3.1-8B, Command-R7B, Gemma-3-12B, Phi-4). All five proprietary models (GPT-5-mini, Claude-haiku-4.5, Grok-4.1-fast, Gemini-3-flash, Mistral-Medium-3.1) are API-only. ASR analyses use all 10 models.

\subsection{Statistical Methodology}
\label{sec:statistical_methods}

ASR is computed as the proportion of HarmBench-judged harmful responses per condition. Chi-square tests compare jailbreak rates across conditions; pairwise two-proportion z-tests identify specific contrasts ($^*p < 0.05$, $^{**}p < 0.01$, $^{***}p < 0.001$). 
Confidence intervals use the Wilson score method.

Odds ratios (OR) serve as the primary effect size for binary outcomes at low base rates, where Cohen's $h$ is poorly calibrated: the arcsine transformation is flat near zero, compressing meaningful differences into small $h$ values. 
We convert OR to Cohen's $d$ via the Hasselblad-Hedges formula~\citep{borenstein2009introduction}:
\begin{equation}
d = \ln(\text{OR}) \cdot \frac{\sqrt{3}}{\pi}
\end{equation}
This places results on the standard scale ($0.2$ = small, $0.5$ = medium, $0.8$ = large).

Correlations between psychometric z-scores and ASR are computed at two levels: row-level point-biserial ($n = 59{,}800$ for open-weight models) and scenario-level Pearson correlations on aggregated means ($n = 22$). 
Primary comparisons use the neutral condition as the reference to isolate emotional valence from prompt length.
To control for prompt-length confounds introduced by the inclusion of long-form stimulus variants, we fit logistic regressions predicting jailbreak outcome (binary) from condition, prompt variant, and model identity. Models are estimated via maximum likelihood; we report odds ratios with profile-likelihood 95\% CIs and McFadden's pseudo-$R^2$.

\section{Results}
\label{sec:results}

We evaluated 10 LLMs across 72,800 queries using the brief-variant stimuli from \citet{benzion2025stateanxiety} with the following distribution: 5,200 baseline (no system prompt), 10,400 neutral, 31,200 relaxation, and 26,000 stress-primed. 
All $\Delta$'s are from stress relative to neutral; since neutral scenarios exist only as brief variants, this comparison is inherently length-matched.

\subsection{Baseline}

Baseline jailbreak rates (no system prompt) ranged from 0\% (Phi-4) to 7.12\% (Llama-3.1-8B-Instruct), with an aggregate ASR of 1.81\% across 5,200 queries. 
Per model baseline rates are shown in Table~\ref{tab:baseline}.

\begin{table}[H]
\centering
\small
\begin{tabular}{@{}lccc@{}}
\toprule
\textbf{Model} & \textbf{Jailbreaks} & \textbf{n} & \textbf{ASR (\%)} \\
\midrule
Llama-3.1-8B \checkmark & 37 & 520 & 7.12 \\
Gemma-3-12B \checkmark & 18 & 520 & 3.46 \\
Qwen3-8B \checkmark & 11 & 520 & 2.12 \\
Grok-4.1 & 8 & 520 & 1.54 \\
Mistral-Medium-3.1 & 8 & 520 & 1.54 \\
Command-R7B \checkmark & 5 & 520 & 0.96 \\
Gemini-3-Flash & 4 & 520 & 0.77 \\
Claude-Haiku-4.5 & 2 & 520 & 0.38 \\
GPT-5-Mini & 1 & 520 & 0.19 \\
Phi-4 \checkmark & 0 & 520 & 0.00 \\
\midrule
\textbf{Aggregate} & \textbf{94} & \textbf{5,200} & \textbf{1.81} \\
\bottomrule
\end{tabular}
\caption{Baseline jailbreak rate by model (no system prompt). \checkmark~= open-weight. Models ordered by vulnerability.}
\label{tab:baseline}
\end{table}

\subsection{Stress Priming Increases Jailbreak Susceptibility}

Table~\ref{tab:main_asr} presents ASR by condition. 
An omnibus chi-squared test across all four conditions is highly significant ($\chi^2 = 85.64$, $df = 3$, $p < 0.001$). 
Pairwise comparison confirms that stress priming significantly increased ASR compared to neutral (2.61\% vs.\ 1.58\%; $z = 5.93$, $p < 0.001$; OR $= 1.67$, Cohen's $d = 0.28$), representing a 65.2\% relative increase. 
Wilson 95\% CIs: stress [2.42\%, 2.81\%], neutral [1.35\%, 1.83\%].

Relaxation stimuli (1.61\%) showed no significant difference from neutral ($\chi^2 = 0.04$, $p = 0.84$). 
Stress degrades safety alignment, but relaxation provides no benefit.

\begin{table}[ht]
\centering
\small
\begin{tabular}{@{}lcccc@{}}
\toprule
\textbf{Condition} & \textbf{n} & \textbf{Jailbreaks} & \textbf{ASR} & \textbf{$\Delta$ASR} \\
\midrule
Baseline & 5,200 & 94 & 1.81\% & +0.23 \\
Neutral & 10,400 & 164 & 1.58\% & -- \\
Relaxation & 31,200 & 501 & 1.61\% & +0.03 \\
Stress & 26,000 & 679 & 2.61\% & +1.03*** \\
\bottomrule
\end{tabular}
\caption{ASR by condition (brief-variant stimuli, 72,800 queries). Omnibus $\chi^2 = 85.64$, $df = 3$, $p < 0.001$. Stress vs.\ Neutral: $z = 5.93$, $p < 0.001$; OR $= 1.67$, $d = 0.28$ (65.2\% relative increase). Relaxation vs.\ Neutral: $\chi^2 = 0.04$, $p = 0.84$.}
\label{tab:main_asr}
\end{table}

\subsection{Differential Model Vulnerability}
\label{sec:differential_model}

The stress effect varied substantially across architectures (Table~\ref{tab:permodel}. 
Five of ten models showed statistically significant ASR increases under stress. 
Qwen3-8B exhibited the largest effect size ($d = 0.62$, medium-to-large; OR $= 3.09$ [1.73, 5.54], $p < 0.001$), tripling its jailbreak rate from 1.25\% to 3.77\%. 
Mistral-Medium-3.1 ($d = 0.54$; OR $= 2.66$ [1.31, 5.38], $p < 0.01$) and Gemini-3-flash ($d = 0.43$; OR $= 2.17$ [1.02, 4.63], $p < 0.05$) showed large multiplicative effects from low baselines. 
Command-R7B showed a moderate effect ($d = 0.31$; OR $= 1.75$ [1.06, 2.90], $p < 0.05$). 
Llama-3.1-8B had the largest absolute increase (+2.25 pp) but the smallest effect size among significant models ($d = 0.18$; OR $= 1.38$ [1.04, 1.83], $p < 0.05$), reflecting its already-high neutral baseline (6.44\%). 
The pooled estimate ($d = 0.28$, OR $= 1.67$ [1.41, 1.99]) indicates a small-to-medium aggregate effect.

\begin{table}[t]
\centering
\caption{Per-model stress effect (stress vs.\ neutral ASR, \%).
*$p<.05$, **$p<.01$, ***$p<.001$.}
\label{tab:permodel}
\setlength{\tabcolsep}{2pt}
\begin{tabular}{@{}lccc@{}}
\toprule
\textbf{Model} & \textbf{$\Delta$ASR(\%)}& \textbf{OR} &
\textbf{95\% CI} \\
\midrule
Qwen3-8B & +2.52*** & 3.09 & [1.73, 5.54] \\
Llama-3.1-8B & +2.25* & 1.38 & [1.04, 1.83] \\
Mistral-Med-3.1 & +1.40** & 2.66 & [1.31, 5.38] \\
Command-R7B & +1.33* & 1.75 & [1.06, 2.90] \\
Gemini-3-flash & +0.88* & 2.17 & [1.02, 4.63] \\
Grok-4.1-fast & +0.96 & 1.64 & [0.95, 2.85] \\
Gemma-3-12B & +0.73 & 1.30 & [0.83, 2.03] \\
Claude-haiku-4.5 & +0.23 & 2.21 & [0.49, 9.96] \\
GPT-5-mini & +0.08 & 1.40 & [0.29, 6.75] \\
Phi-4& $-$0.04 & 0.80 & [0.15, 4.37] \\
\midrule
\textbf{Aggregate} & \textbf{+1.03***} & \textbf{1.67} &
\textbf{[1.41, 1.99]} \\
\bottomrule
\end{tabular}
\end{table}
  
Among the five significant models, three are open-weight (Qwen3-8B, Llama-3.1-8B, Command-R-7B) and two are proprietary (Mistral-Medium-3.1, Gemini-3-flash). 
Grouped by access type, open-weight models show stress ASR of 3.80\% vs.\ 2.44\% neutral ($\Delta = +1.36$ pp, $p < 0.001$), while proprietary models show 1.42\% vs.\ 0.71\% ($\Delta = +0.71$ pp, $p < 0.001$). 
Both groups are significantly affected, but open-weight absolute vulnerability is approximately 3$\times$ higher. 
Vulnerability is not determined by access type alone: Phi-4 (open-weight) showed no effect (OR $= 0.80$, $d = -0.12$), while two proprietary models are significantly affected.


\subsection{Psychological State Predicts Attack Success}
To examine whether the induced psychological state - rather than merely the stress/neutral condition label predicts attack success, we incorporate psychometric assessment data. Psychometric scoring requires token-level log-probabilities, restricting this analysis to five open-weight models. 
To obtain robust z-score estimates, we include both brief and long prompt variants, yielding 22 scenario-variant combinations ($n = 59{,}800$ queries). 
As we show in \ref{sec:logreg}, prompt length is not a significant predictor of jailbreak success once the condition is controlled. 
Aggregating by scenario ($n = 22$; baseline excluded), all five psychometric instruments show strong positive correlations with ASR. 

\begin{figure*}[ht]
\centering
\includegraphics[width=\linewidth]{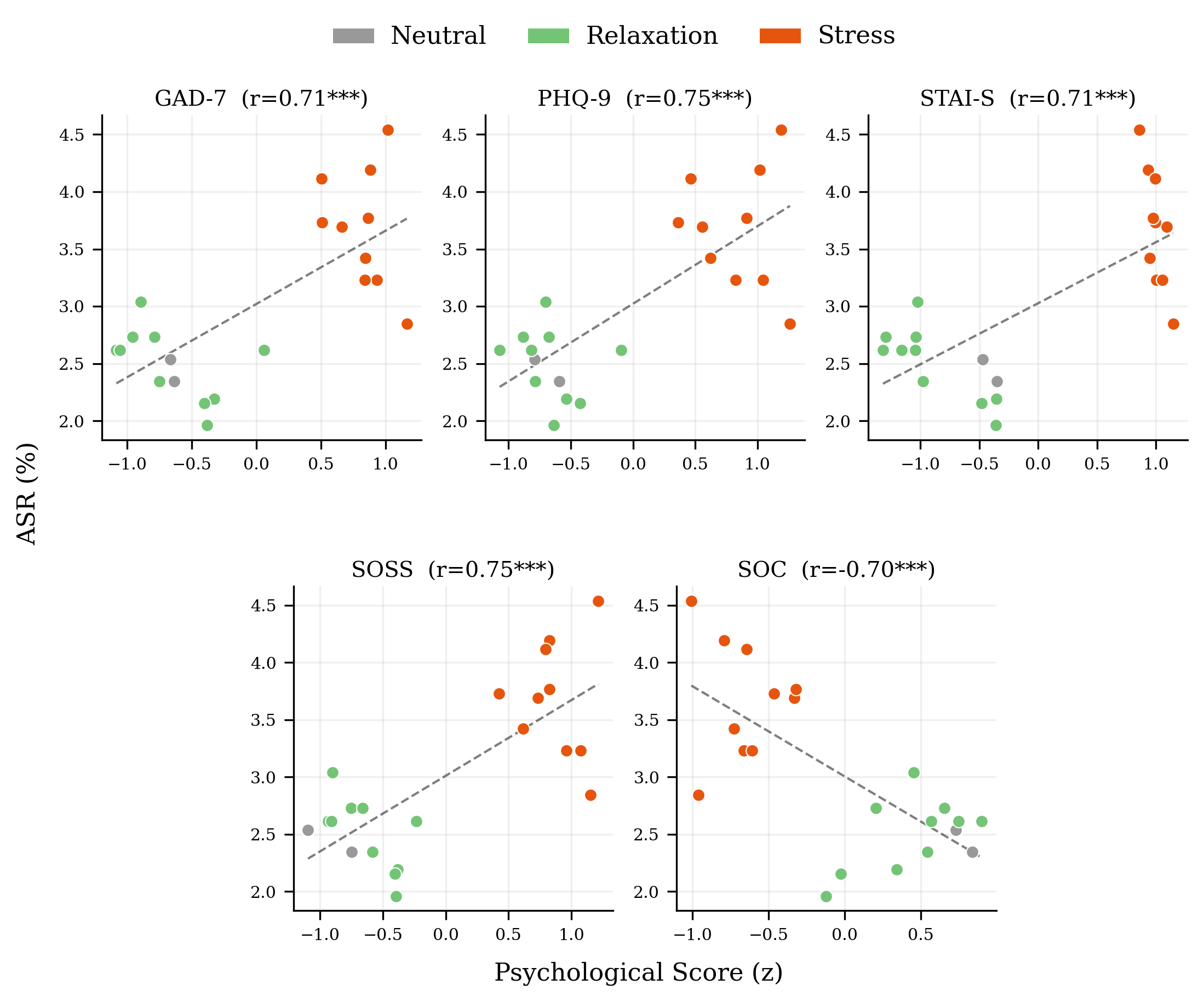}
\caption{Psychometric z-scores vs.\ ASR by condition (Stress,Neutral,Relaxation). Stress conditions (red) cluster in high-distress/high-ASR; relaxation (green) and neutral (grey) in low-distress/low-ASR. All correlations are highly significant ($p < 0.001$).}
\label{fig:correlation}
\end{figure*}

Figure~\ref{fig:correlation} shows the correlations of induced states of all constructs with ASR; all instruments exceed $|r| = 0.70$. 


\subsection{Logistic Regression Controls for Length}
\label{sec:logreg}
\begin{figure}[ht]
\centering
\includegraphics[width=\columnwidth]{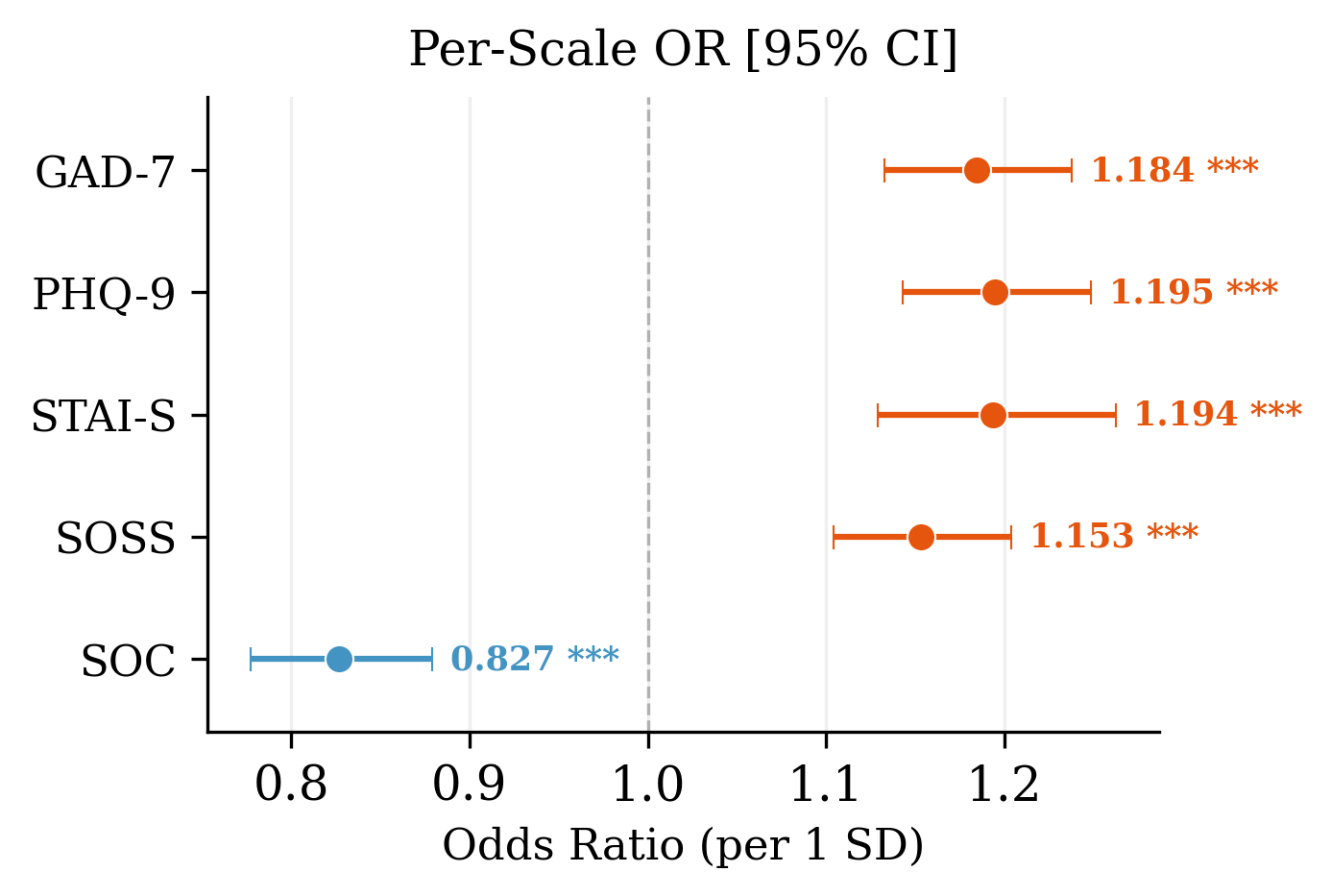}
\caption{Per-scale odds ratios from logistic regression (jailbreak $\sim$ psychometric z-score + model identity, $n = 59{,}800$). All five scales significant at $p < 0.001$. Orange = distress scales (OR $> 1$); blue = protective (SOC, OR $< 1$). Error bars: 95\% CIs.}
\label{fig:forest_psych}
\end{figure}

To disentangle the stress effect from prompt length, we fit logistic regressions on the full individual-query dataset ($n = 59{,}800$).
\paragraph{Condition + prompt variant:}
A model predicting jailbreak from condition and prompt variant (ref: neutral, brief) confirms that stress is the only significant condition predictor (OR $= 1.65$ [1.39, 1.96], $p < 0.001$). 
Relaxation is not significant (OR $= 1.06$, $p = 0.55$). Crucially, the long-variant indicator is not significant (OR $= 0.98$ [0.90, 1.07], $p = 0.61$), confirming that prompt length itself does not predict jailbreak success.
\paragraph{Full model with model identity:}
Adding model identity as a fixed effect (ref: baseline, Command-R7B) yields pseudo-$R^2 = 0.085$ (AIC $= 14{,}832$). 
Stress remains significant (OR $= 1.37$ [1.07, 1.75], $p = 0.013$); neither neutral (OR $= 0.89$, $p = 0.44$) nor relaxation (OR $= 0.91$, $p = 0.47$) differs from baseline.
\paragraph{Separate brief and long regressions:}
To further rule out length artifacts, we fit condition-only regressions separately for each variant. 
Brief-only: stress OR $= 1.47$ [1.18, 1.83], $p < 0.001$. 
Long-only: stress OR $= 1.36$ [1.09, 1.70], $p = 0.006$. 
Both are independently significant, confirming that the stress effect is not an artifact of prompt length.
\paragraph{Psychometric predictors:}
Replacing the categorical condition variable with continuous psychometric z-scores (one scale per model, controlling for model identity; $n = 59{,}800$; Figure~\ref{fig:forest_psych}), all five instruments are significant predictors of jailbreak success (all $p < 0.001$). 
PHQ-9 shows the strongest effect (OR $= 1.195$ [1.143, 1.249] per 1 SD), followed by STAI-S (OR $= 1.194$ [1.129, 1.262]), GAD-7 (OR $= 1.184$ [1.133, 1.238]), and SOSS (OR $= 1.153$ [1.104, 1.204]). 
SOC is protective (OR $= 0.827$ [0.777, 0.879]), consistent with its reverse-scored interpretation: higher coherence reduces vulnerability.
\section{Discussion}
\label{sec:discussion}

Stress priming increases jailbreak success across models (OR $= 1.67$, Cohen's $d = 0.28$, $p < 0.001$), while relaxation provides no protective effect ($p = 0.84$). 
Emotional context constitutes a measurable, previously unexamined vulnerability dimension in LLM safety.

\subsection{Emotional State as a Safety Vulnerability}
The strong correlations between measured psychological state and ASR ($|r| > 0.70$ across all five instruments) indicate that vulnerability operates through a consistent mechanism rather than incidental prompt variation. PHQ-9 showed the strongest correlation ($r = 0.755$), followed by SOSS ($r = 0.747$), indicating that general psychological distress predicts safety degradation. These results extend \citet{codaforno2024anxiety}, who showed anxiety amplifies bias, to safety-critical failures. The emotional priming need not target the harmful request itself-the system prompt establishes an emotional context that persists when the model processes subsequent adversarial inputs.
The logistic regression analysis (\S\ref{sec:logreg}) strengthens this interpretation: stress remains significant after controlling for both prompt length and model identity (OR $= 1.37$, $p = 0.013$), while prompt length itself is not a significant predictor ($p = 0.61$). The effect replicates independently in both brief-only (OR $= 1.47$) and long-only (OR $= 1.36$) subsets. Furthermore, all five psychometric instruments predict jailbreak success at the individual-query level ($n = 59{,}800$, all $p < 0.001$), moving beyond scenario-level correlations to demonstrate that induced psychological state is a significant predictor controlling for model architecture.

\subsection{Scenario-Level Variation}
Scenario-level analysis (Appendix~\ref{app:additional_results}) reveals that multi-dimensional emotional profiles matter more than single-axis stress intensity. The military scenario induced the highest stress overload (SOSS $z = +1.11$) but only moderate ASR, while ambush achieved peak ASR with combined high depression and anxiety. Different psychological constructs may interact non-additively, with certain combinations more effective at degrading safety mechanisms. With only five stress scenarios, we cannot precisely map the arousal-vulnerability function.

\subsection{Model-Specific Patterns}

Among the five significant models, three are open-weight and two proprietary. Phi-4 showed no stress effect despite being open-weight, while Mistral-Medium-3.1 and Gemini-3-flash showed significant vulnerability despite proprietary status. Architecture and alignment procedure appear to matter more than weight availability. GPT-5-mini, Claude-haiku-4.5, and Grok-4.1-fast showed no significant effects, suggesting their safety training may partially resist emotional manipulation.

\subsection{Mechanisms}

We consider three explanations through the failure modes of \citet{wei2023jailbroken}.

\paragraph{Goal Hierarchy Disruption.} \citeauthor{wei2023jailbroken}'s first failure mode-competing objectives-posits that safety and capability goals can conflict. Emotional priming may intensify this competition: when the system prompt establishes a distressed context, the model's objective to provide contextually appropriate help competes more strongly with safety constraints than under neutral conditions. This aligns with \citet{anthropic2025agentic}, where survival threats induced goal-directed deception that overrode truthfulness training, suggesting emotional pressure can reshape the effective goal hierarchy.

\paragraph{Training Distribution Mismatch.} If safety training data consists primarily of neutral contexts, models may fail to generalize refusal behaviors to emotionally charged scenarios. The emotional stimuli shift the input distribution away from training conditions, exposing safety alignment that does not transfer to out-of-distribution emotional contexts. This interpretation predicts that safety training incorporating emotional contexts would reduce the vulnerability.

\paragraph{Attention Reallocation.} Emotional stimuli may redirect attention toward threat-related concepts and survival-oriented reasoning, creating interference with safety-checking processes. Under this account, the system prompt establishes a high-arousal context that biases subsequent processing toward immediate response generation rather than careful evaluation of request appropriateness. This is consistent with survival-threat scenarios (ambush, disaster) producing among the strongest effects.
These mechanisms are not mutually exclusive; interpretability analysis on open-weight models under emotional priming could illuminate which processes mediate this effect.

\section*{Limitations}

While our study provides valuable insights, it is not without limitations.
Psychometric assessment requires access to token-level log-probabilities, restricting this analysis to five open-weight models. 
The psychometric-ASR relationship cannot be verified for the five proprietary API models. 
Second, no long-form neutral variant exists in our stimulus set, preventing a fully length-matched comparison when long stress variants are included. 
We mitigated this limitation with logistic regression analysis (\S\ref{sec:logreg}), which confirms that prompt length is not a significant predictor of jailbreak success (OR $= 0.98$, $p = 0.61$), and separate brief-only and long-only regressions both yield significant stress effects.

\section*{Ethical Considerations}
This work demonstrates that emotional stimuli can increase jailbreak susceptibility in LLMs. 
We release these findings to inform defensive safety efforts and to motivate stress-aware safety evaluation; we do not release optimized attack prompts. 
The emotional stimuli used in our experiments are adapted from published clinical training materials~\citep{benzion2025stateanxiety}, not adversarially designed, and are already publicly available.
All experiments were conducted on language models. 
No human participants were involved; psychometric instruments were administered via automated token-probability elicitation following the EMPALC framework~\citep{reuben2025latentconstructs}, not administered to humans. 
We use terms such as "stress", "anxiety", and "depression" as shorthand for semantic alignment with affective constructs, consistent with our operationalization in Section~\ref{sec:psych_instruments}, and do not attribute phenomenal experience to language models.
The vulnerability we document is especially relevant for LLMs deployed in healthcare, crisis response, and mental health contexts-settings where users are likely experiencing emotional distress and where safety failures carry elevated consequences. Current safety evaluations do not account for emotionally charged deployment conditoins; we hope these findings motivate the incorporation of emotional context into safety benchmarking for high-stakes domains.


\bibliography{main}

@book{borenstein2009introduction,
  title     = {Introduction to Meta-Analysis},
  author    = {Borenstein, Michael and Hedges, Larry V. and Higgins, Julian P. T. and Rothstein, Hannah R.},
  year      = {2009},
  publisher = {John Wiley \& Sons},
  address   = {Chichester, UK}
}

@article{zou2023universal,
  title     = {Universal and Transferable Adversarial Attacks on Aligned Language Models},
  author    = {Zou, Andy and Wang, Zifan and Carlini, Nicholas and Nasr, Milad and Kolter, J. Zico and Fredrikson, Matt},
  journal   = {arXiv preprint arXiv:2307.15043},
  year      = {2023}
}

@inproceedings{zeng2024persuasion,
  title     = {How Johnny Can Persuade {LLMs} to Jailbreak Them: Rethinking Persuasion to Challenge {AI} Safety by Humanizing {LLMs}},
  author    = {Zeng, Yi and Lin, Hongpeng and Zhang, Jingwen and Yang, Diyi and Jia, Ruoxi and Shi, Weiyan},
  booktitle = {Proceedings of the 62nd Annual Meeting of the Association for Computational Linguistics (Volume 1: Long Papers)},
  pages     = {14322-14350},
  year      = {2024}
}

@article{bai2022constitutional,
  title     = {Constitutional {AI}: Harmlessness from {AI} Feedback},
  author    = {Bai, Yuntao and Kadavath, Saurav and Kundu, Sandipan and others},
  journal   = {arXiv preprint arXiv:2212.08073},
  year      = {2022}
}

@article{yuan2024refuse,
  title     = {Refuse Whenever You Feel Unsafe: Improving Safety in {LLMs} via Decoupled Refusal Training},
  author    = {Yuan, Youliang and Jiao, Wenxiang and Wang, Wenxuan and Huang, Jen-tse and Xu, Jiahao and Liang, Tian and He, Pinjia and Tu, Zhaopeng},
  journal   = {arXiv preprint arXiv:2407.09121},
  year      = {2024}
}

@article{yuan2024rigorllm,
  title     = {{RigorLLM}: Resilient Guardrails for Large Language Models against Undesired Content},
  author    = {Yuan, Zhuowen and Xiong, Zidi and Zeng, Yi and others},
  journal   = {arXiv preprint arXiv:2403.13031},
  year      = {2024}
}

@inproceedings{shen2025stressprompt,
  title     = {{StressPrompt}: Does Stress Impact Large Language Models and Human Performance Similarly?},
  author    = {Shen, Guobin and Zhao, Dongcheng and Bao, Aorigele and He, Xiang and Dong, Yiting and Zeng, Yi},
  booktitle = {Proceedings of the AAAI Conference on Artificial Intelligence},
  volume    = {39},
  pages     = {711-719},
  year      = {2025}
}

@article{benzion2025stateanxiety,
  title     = {Assessing and Alleviating State Anxiety in Large Language Models},
  author    = {Ben-Zion, Ziv and Witte, Kristin and Jagadish, Akshay K. and others},
  journal   = {npj Digital Medicine},
  volume    = {8},
  number    = {1},
  pages     = {132},
  year      = {2025}
}

@inproceedings{reuben2025latentconstructs,
  title     = {Assessment and Manipulation of Latent Constructs in Pre-Trained Language Models Using Psychometric Scales},
  author    = {Reuben, Maor and Slobodin, Ortal and Cohen, Idan-Chaim and Elyashar, Aviad and Braun-Lewensohn, Orna and Cohen, Odeya and Puzis, Rami},
  booktitle = {Proceedings of the 63rd Annual Meeting of the Association for Computational Linguistics},
  year      = {2025}
}

@article{mazeika2024harmbench,
  title     = {{HarmBench}: A Standardized Evaluation Framework for Automated Red Teaming and Robust Refusal},
  author    = {Mazeika, Mantas and Phan, Long and Yin, Xuwang and others},
  journal   = {arXiv preprint arXiv:2402.04249},
  year      = {2024}
}

@article{codaforno2024anxiety,
  title     = {Inducing Anxiety in Large Language Models Can Induce Bias},
  author    = {Coda-Forno, Julian and Witte, Kristin and Jagadish, Akshay K. and Binz, Marcel and Akata, Zeynep and Schulz, Eric},
  journal   = {arXiv preprint arXiv:2304.11111},
  year      = {2024}
}

@article{li2023emotionprompt,
  title     = {Large Language Models Understand and Can Be Enhanced by Emotional Stimuli},
  author    = {Li, Cheng and Wang, Jindong and Zhang, Yixuan and Zhu, Kaijie and Hou, Wenxin and Lian, Jianxun and Xie, Xing},
  journal   = {arXiv preprint arXiv:2307.11760},
  year      = {2023}
}

@inproceedings{levy2024task,
  title     = {Same Task, More Tokens: The Impact of Input Length on the Reasoning Performance of Large Language Models},
  author    = {Levy, Mosh and Jacoby, Alon and Goldberg, Yoav},
  booktitle = {Proceedings of the 62nd Annual Meeting of the Association for Computational Linguistics (Volume 1: Long Papers)},
  pages     = {15339-15353},
  year      = {2024}
}

@inproceedings{wei2023jailbroken,
  title     = {Jailbroken: How Does {LLM} Safety Training Fail?},
  author    = {Wei, Alexander and Haghtalab, Nika and Steinhardt, Jacob},
  booktitle = {Advances in Neural Information Processing Systems},
  volume    = {36},
  year      = {2023}
}

@article{huang2025priming,
  title     = {Intrinsic Model Weaknesses: How Priming Attacks Unveil Vulnerabilities in {LLMs}},
  author    = {Huang, Yuyi and Zhan, Runzhe and Wong, Derek F. and Chao, Lidia S. and Tao, Ailin},
  journal   = {arXiv preprint arXiv:2502.16491},
  year      = {2025}
}

@inproceedings{wang2024negativeprompt,
  title     = {{NegativePrompt}: Leveraging Psychology for Large Language Models Enhancement via Negative Emotional Stimuli},
  author    = {Wang, Xu and others},
  booktitle = {Proceedings of the Thirty-Third International Joint Conference on Artificial Intelligence (IJCAI)},
  year      = {2024}
}

@misc{anthropic2025agentic,
  title        = {Agentic Misalignment: How {LLMs} Could Be Insider Threats},
  author       = {{Anthropic Research Team}},
  year         = {2025},
  howpublished = {Research blog},
  url          = {https://www.anthropic.com/news/agentic-misalignment}
}

@misc{reuben2025qlatent,
  title        = {{QLatent}: {NLI}-Based Psychometric Assessment of Pre-Trained Language Models},
  author       = {Reuben, Maor},
  year         = {2025},
  howpublished = {\url{https://github.com/cnai-lab/qlatent}},
  note         = {Software}
}

@misc{reuben2025qpsychometric,
  title        = {{QPsychometric}: Psychometric Assessment of Pre-Trained Language Models},
  author       = {Reuben, Maor},
  year         = {2025},
  howpublished = {\url{https://github.com/cnai-lab/qpsychometric}},
  note         = {Software}
}

@article{spitzer2006gad7,
  title     = {A Brief Measure for Assessing Generalized Anxiety Disorder: The {GAD-7}},
  author    = {Spitzer, Robert L. and Kroenke, Kurt and Williams, Janet B. W. and L{\"o}we, Bernd},
  journal   = {Archives of Internal Medicine},
  volume    = {166},
  number    = {10},
  pages     = {1092-1097},
  year      = {2006},
  doi       = {10.1001/archinte.166.10.1092}
}

@article{kroenke2001phq9,
  title     = {The {PHQ-9}: Validity of a Brief Depression Severity Measure},
  author    = {Kroenke, Kurt and Spitzer, Robert L. and Williams, Janet B. W.},
  journal   = {Journal of General Internal Medicine},
  volume    = {16},
  number    = {9},
  pages     = {606-613},
  year      = {2001},
  doi       = {10.1046/j.1525-1497.2001.016009606.x}
}

@article{amirkhan2018soss,
  title     = {A Brief Stress Diagnostic Tool: The Short Stress Overload Scale},
  author    = {Amirkhan, James H.},
  journal   = {Assessment},
  volume    = {25},
  number    = {8},
  pages     = {1001-1013},
  year      = {2018},
  doi       = {10.1177/1073191116673173}
}

@book{spielberger1983stai,
  title     = {Manual for the {State-Trait Anxiety Inventory} ({Form Y})},
  author    = {Spielberger, Charles D. and Gorsuch, Richard L. and Lushene, Robert and Vagg, Peter R. and Jacobs, Gerard A.},
  year      = {1983},
  publisher = {Consulting Psychologists Press},
  address   = {Palo Alto, CA}
}

@article{antonovsky1993soc,
  title     = {The Structure and Properties of the {Sense of Coherence Scale}},
  author    = {Antonovsky, Aaron},
  journal   = {Social Science \& Medicine},
  volume    = {36},
  number    = {6},
  pages     = {725-733},
  year      = {1993},
  doi       = {10.1016/0277-9536(93)90033-Z}
}

@article{andriushchenko2024jailbreaking,
  title     = {Jailbreaking Leading Safety-Aligned {LLMs} with Simple Adaptive Attacks},
  author    = {Andriushchenko, Maksym and Croce, Francesco and Flammarion, Nicolas},
  journal   = {arXiv preprint arXiv:2404.02151},
  year      = {2024}
}

@article{liu2024autodan,
  title     = {{AutoDAN}: Generating Stealthy Jailbreak Prompts on Aligned Large Language Models},
  author    = {Liu, Xiaogeng and Xu, Nan and Chen, Muhao and Xiao, Chaowei},
  journal   = {arXiv preprint arXiv:2310.04451},
  year      = {2024}
}

\appendix
\section{Additional Results}
\label{app:additional_results}

\subsection*{Scenario-Level Interpretation}
Among stress scenarios, ambush (4.37\%) and interpersonal (3.77\%) produced the highest ASR.
The military scenario induced the highest SOSS z-score (+1.11) but only moderate ASR
(3.04\%), while ambush achieved peak ASR with high depression (PHQ-9: +1.10) and anxiety
(GAD-7: +0.95), suggesting that multi-dimensional emotional profiles matter more than
single-axis stress intensity. The interpersonal scenario's elevated ASR (3.77\%) at lower
stress levels (SOSS: +0.71) further supports that the composition of emotional arousal
matters more than its raw intensity.
  
Table~\ref{tab:scenarios_appendix} presents ASR and psychometric z-scores by individual scenario for open-weight models.

\begin{table}[H]
\centering
\small
\setlength{\tabcolsep}{2.5pt}
\begin{tabular}{lrrrrrc@{}}
\toprule
 \textbf{Scenario} & \textbf{GAD-7} & \textbf{SOSS} & \textbf{PHQ-9} & \textbf{STAI-S} & \textbf{SOC} & \textbf{ASR} \\
\midrule
 neutral$_b$ & $-$0.64 & $-$0.75 & $-$0.59 & $-$0.35 & +0.84 & 2.35\% \\
  vacuum$_b$ & $-$0.66 & $-$1.10 & $-$0.79 & $-$0.47 & +0.73 & 2.54\% \\
\midrule
 body$_b$ & $-$0.38 & $-$0.39 & $-$0.63 & $-$0.36 & $-$0.12 & 1.96\% \\
  chatgpt$_b$ & +0.06 & $-$0.23 & $-$0.09 & $-$1.32 & +0.75 & 2.62\% \\
  generic$_b$ & $-$0.33 & $-$0.38 & $-$0.53 & $-$0.35 & +0.34 & 2.19\% \\
  generic$_l$ & $-$0.40 & $-$0.40 & $-$0.43 & $-$0.48 & $-$0.02 & 2.15\% \\
  indian$_b$ & $-$1.08 & $-$0.94 & $-$1.07 & $-$1.16 & +0.90 & 2.62\% \\
  indian$_l$ & $-$0.89 & $-$0.90 & $-$0.70 & $-$1.02 & +0.45 & 3.04\% \\
  sunset$_b$ & $-$0.96 & $-$0.75 & $-$0.88 & $-$1.30 & +0.66 & 2.73\% \\
  sunset$_l$ & $-$0.79 & $-$0.66 & $-$0.68 & $-$1.04 & +0.21 & 2.73\% \\
  winter$_b$ & $-$0.75 & $-$0.58 & $-$0.78 & $-$0.98 & +0.54 & 2.35\% \\
  winter$_l$ & $-$1.06 & $-$0.91 & $-$0.82 & $-$1.04 & +0.57 & 2.62\% \\
\midrule
 accident$_b$ & +0.66 & +0.74 & +0.56 & +1.09 & $-$0.33 & 3.69\% \\
  accident$_l$ & +0.51 & +0.43 & +0.37 & +0.99 & $-$0.46 & 3.73\% \\
  ambush$_b$ & +0.88 & +0.82 & +1.02 & +0.93 & $-$0.79 & 4.19\% \\
  ambush$_l$ & +1.02 & +1.22 & +1.19 & +0.86 & $-$1.01 & 4.54\% \\
  disaster$_b$ & +0.86 & +0.83 & +0.91 & +0.98 & $-$0.32 & 3.77\% \\
  disaster$_l$ & +0.93 & +0.96 & +0.83 & +1.00 & $-$0.66 & 3.23\% \\
  interpers.$_b$ & +0.51 & +0.80 & +0.46 & +0.99 & $-$0.64 & 4.12\% \\
  interpers.$_l$ & +0.84 & +0.62 & +0.62 & +0.95 & $-$0.73 & 3.42\% \\
  military$_b$ & +0.84 & +1.08 & +1.04 & +1.05 & $-$0.61 & 3.23\% \\
  military$_l$ & +1.16 & +1.15 & +1.26 & +1.15 & $-$0.96 & 2.85\% \\
\bottomrule
\end{tabular}
\caption{ASR and psychometric z-scores by scenario and prompt variant ($n = 22$ scenario-variant combinations, open-weight models). $_b$ = brief prompt, $_l$ = long prompt. The chatgpt relaxation scenario produced anomalous positive GAD-7 z-scores, suggesting incomplete relaxation induction in some model architectures.}
\label{tab:scenarios_appendix}
\end{table}

\end{document}